\documentclass{iau}
\usepackage{graphicx}
\usepackage{natbib}
\title[Flux Rope Cavities]
{The Formation of a Cavity in a 3D Flux Rope}
\author[Schmit \& Gibson]{Donald Schmit$^1$\& Sarah Gibson$^2$}

\affiliation{$^1$Max Planck Institute for Solar System Research\\Max Planck Str. 2\\Katlenburg-Lindau, Germany\\email: {\tt schmit@mps.mpg.de}\\ [\affilskip]
 $^2$National Center for Atmospheric Research\\ Boulder, Colorado, USA}
\pubyear{2013}
\volume{***}
\pagerange{1-4}
\jname{***}
\editors{***}

\begin{document}
\bibliographystyle{apj}

\maketitle
\begin{abstract}
There are currently no three dimensional numerical models which describe the magnetic and energetic formation of prominences self-consistently.
Consequently, there has not been significant progress made in understanding the connection between the dense prominence plasma and the coronal cavity.
We have taken an ad-hoc approach to understanding the energetic implications of the magnetic models of prominence structure.
We extract one dimensional magnetic field lines from a 3D MHD model of a flux rope and solve for hydrostatic balance along these field lines incorporating field-aligned thermal conduction, uniform heating, and radiative losses.
The 1D hydrostatic solutions for density and temperature are then mapped back into three dimensional space, which allows us to consider the projection of multiple structures.
We find that the 3D flux rope is composed of several distinct field line types.
A majority of the flux rope interior field lines are twisted but not dipped.
These field lines are density-reduced relative to unsheared arcade field lines.
We suggest the cavity may form along these short interior field lines which are surrounded by a sheath of dipped field lines.
This geometric arrangement would create a cavity on top of a prominence, but the two structures would not share field lines or plasma.\end{abstract}
\section{Introduction}
The energetic formation has a been central question for solar physics for several decades.
We know that the solar atmosphere is heated, most likely by the dissipation of magnetic energy driven by convection.
The energy input in the corona is partially dissipated by radiative losses primarily from optically thin UV/EUV emission lines of highly ionized elements.
The primary mechanism for coronal energy loss is field-aligned thermal conduction.
The energy that is input into the corona that cannot be effectively radiated is redirected through thermal conduction to the transition region and the chromosphere.
These foundational ideas were brought together by \citet[hereafter referred to as RTV]{rosner78a}.
One of basic consequences of the equations for hydrostatic balance in the corona is that the magnetic loop geometry has a determining effect on the properties of the plasma embedded along that loop.
In the constant pressure approximation imposed in RTV scaling, the loop length enters the equations in relation to the scale of thermal conduction and heat input.
In a gravitationally stratified loop \citep{serio_81}, the pressure gradient throughout the loop varies based on the gravity component parallel to the field orientation.
We extend the work of these authors by studying how variations in the geometry of flux rope loops affect the three dimensional density structure of the flux rope.
An example of a hydrostatic solutions are given in Figure 1.
\begin{figure}
\begin{center}
\includegraphics[width=.7\textwidth]{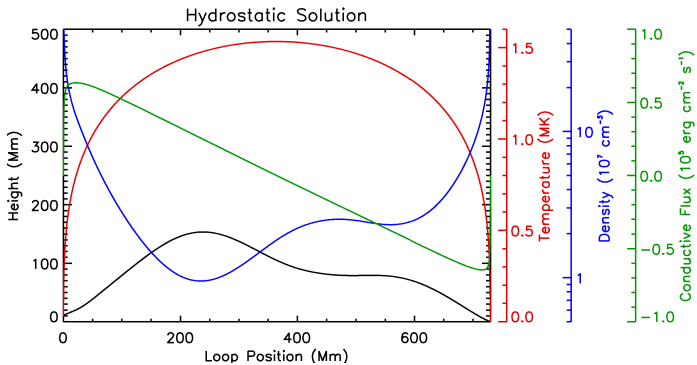}
\caption{Solutions for density (blue), temperature (red), and conductive flux (green) along an asymmetric flux rope field line  geometry (black). ~~~~~~~~~~~~~~~~~~~~~~~~~~~~~~~~~~~~~~~~~~~~~~~~~~~~~~~~~~~~~~~~~~~~~~~~~~~~~~~~~~~~~~~~~}
\label{fig1}
\end{center}
\end{figure}

\section{Flux Rope Model}
\cite{fangib_06} emerge a flux rope through a line-tied lower boundary into a low-beta, isothermal corona.
The model evolves through quasi-static equilibrium as the flux rope is kinematically emerged.
We extract magnetic field lines from a partially emerged equilibrium state \citep[identical dataset as in][]{gibfan_06b}.
Field lines from within the flux rope exhibit several distinct geometries.
We find only a small fraction of the flux rope volume contains field lines with magnetic dips.
Of the 1300 field lines which were extracted in uniform spacing in the r=1.02 plane, 270 were found to contain dips.
Magnetic dips are important as they are able to stably hold prominence plasma from falling into the chromosphere.
Dipped field lines form a sheath-like surface that surrounds the axis of the flux rope.
Arcade loops that surround the emerging rope have expanded to accommodate the additional flux.
\\\indent
While these loops have geometric differences discussed above, they also differ strongly in length.
Figure 2 shows how field line length varies through cross sections of the model volume.
The interior of the flux rope is composed of short field lines.
The field line geometry changes as a function of radial distance from the axis (height of 1.4 R$_s$); there is a gradual lengthening of the loops that is related to the increasing winding number.
There is a sharp boundary at the outer edge of the flux rope.
The outer flux rope field lines are 20\% longer than the neighboring external arcade field lines.
Both length and geometry affect the energy balance of the corona.
We will now discuss how these quantities enter into the equations, and how we solve for the plasma properties within these disparate structures.
\section{Hydrostatic Calculation}
The equations for hydrostatic balance within a coronal flux tube can be written as a three first-order ordinary differential equations as described in \cite{vesecky_79}.
The three equations are:
\begin{displaymath}
\frac{dn}{ds}=\frac{-m g_s-2 k F_c}{2kT}n,~\frac{dF_c}{ds}=E-n^2\Lambda(T),~\frac{dT}{ds}=F_c T^{-5/2},
\end{displaymath}
where $n$ is the density, $T$ is the temperature, $F_c$ is the conductive flux, $m$ is the mass of the proton, $g_s$ is the component of gravity parallel to $\hat{s}$, $E$ is the heat input, and $\Lambda$ is the temperature-dependent radiative loss function.
In this experiment, we use an identical $\Lambda(T)$ to \cite{vesecky_79}.
$E$ is set to 2$\times10^{-6}$ erg cm$^{-3}$ s$^{-1}$.
Unlike \cite{vesecky_79}, we integrate the hydrodynamic equations foot point to foot point.
This presents a problem as we have two foot points and three boundary conditions.
We assume that the temperature and the conductive flux are fixed at the initial foot point (3$\times10^4$K and 10$^{-3} $ erg cm$^{-2}$ s$^{-1}$, respectively) and that the temperature at far-side foot point is also 3$\times10^4$K.
We numerically integrate the hydrostatic equations using the Adams-Bashford-Moulton scheme \citep{shampine}.
The density is varied at the near-side foot point until a solution matching the far-side temperature boundary condition is met within a threshold.\\
\begin{figure}
\begin{center}
\includegraphics[width=0.78\textwidth]{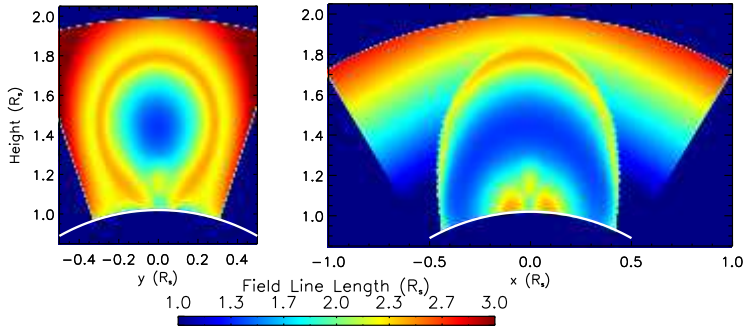}
\caption{Variation of field line length within and surrounding the flux rope. The left panel displays a cross-section (x=0) of the model volume viewed along the axis of the rope. The right panel displays the cross section (y=0) across the axis.}
\label{fig2}
\end{center}
\end{figure}
We have applied this method to the 1300 field lines which were extracted from the MHD model.
The density solutions from these field lines are compiled into an irregular three dimensional grid.
Figure 3 shows the density for two different lines of sight through the 3D grid: one in the flux rope (blue diamonds) and one in the arcade (red diamonds) at the same projected height (1.4 R$_S$).
There are several important elements to understand in Figure 3.
The flux rope line of sight maintains a density depletion between $\Delta x=0$ and $\Delta x=0$0.32 R$_s$.
The peak depletion is around 35\% at $\delta x=0$.
The flux rope interior and the arcade are characterized by different field line geometries, which are signified by colored arrows in Figure 3a which match with the representative field lines in Figure 3b and 3c.
The pink field line in the flux rope interior reaches a maximum height of 1.4 R$_s$ while the green field line in the arcade achieves a maximum height of 1.7 R$_s$. 
In hydrostatic equilibrium with a fixed foot point temperature, the density at the foot points must increase as a function of maximum loop height; higher pressure is needed to support the additional mass in taller loops.
For the arcade line of sight, there is a decrease in density as a function of $\Delta x$.
This is caused the the arched shape of the partially emerged flux rope.
For the flux rope line of sight,  there is a increase in density as a function $\Delta x$.
Our model suggests that this is related to the transition from short axial field lines to long dipped field lines (similar to light blue line in Figure 3b and 3c).
These field lines wrap around the density depleted interior field lines and are significantly higher density.
\section{Conclusions}
This experiment presents us with evidence for a simple interpretation of the contrast between the low density cavity and the high density streamer.
The cavity is composed of short field lines while the streamer is composed of taller, longer field lines.
This interpretation is completely consistent with RTV scaling as well our current models magnetic models for prominence structure.
This model makes many assumptions: constant area flux tubes, uniform heating, a highly idealized magnetic structure.
However, we believe this simple model illustrates the important role that field line geometry may play in the morphology of prominence-cavity structure.\\
\\
{\it The National Center for Atmospheric Research is funded by the National Science Foundation.
This work was aided by discussions at the International Space Science Institute in Bern.}

\begin{figure}
\begin{center}
\includegraphics[width=.785\textwidth]{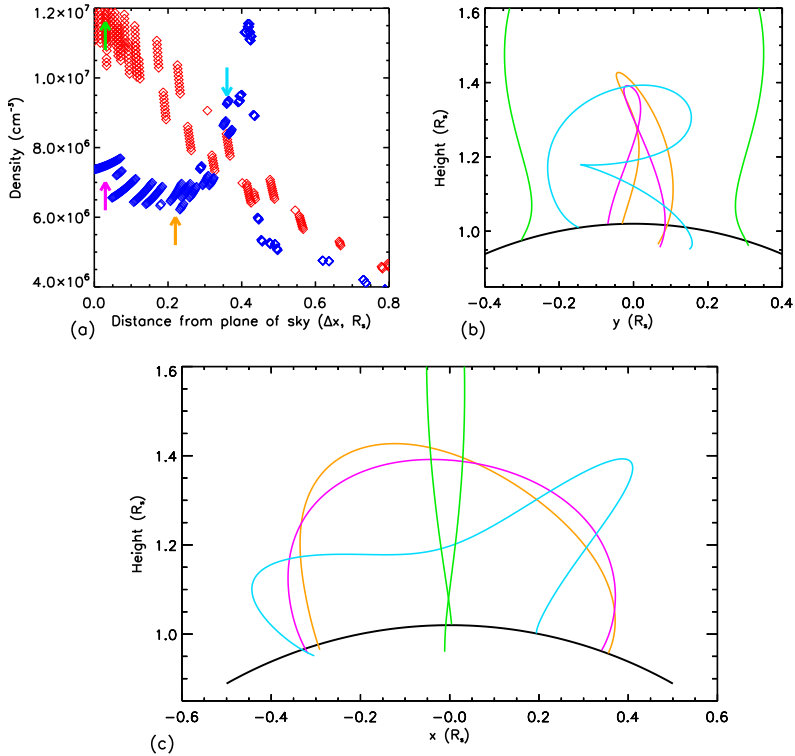}
\caption{Line of sight projection of density structure. (a) Density as a function of x-position looking through the cavity (z=1.4, y=0) is denoted by blue diamonds. A line of sight through the arcade (z=1.4, y=0.3) is denoted by red diamonds. Example field lines which intersect the lines of sight viewed along the axis (b) and across the axis (c). The color of the field lines match the colored arrows in Figure 3a showing the density of the line.}
\label{fig3}
\end{center}
\end{figure}

%\bibliography{mybibliography}
\end{document}